\documentclass{emulateapj}
\usepackage{threeparttable}
\usepackage{natbib}
\usepackage[bookmarks=false, colorlinks=true, citecolor=blue,
  backref=true]{hyperref}              
\hypersetup{
  colorlinks,
  citecolor=blue,
  linkcolor=red,
  urlcolor=blue,
}

\DeclareRobustCommand{\ion}[2]{%
\relax\ifmmode
\ifx\testbx\f@series
{\mathbf{#1\,\mathsc{#2}}}\else
{\mathrm{#1\,\mathsc{#2}}}\fi
\else\textup{#1\,{\mdseries\textsc{#2}}}%
\fi}

\shorttitle{AGC 111629}
\shortauthors{Cao et al.}

\begin{document}
\slugcomment{{\bf Accepted for publication in AJ}}

\title{Ionized gas components in low surface brightness galaxy AGC 111629}
\author{Tian-Wen Cao\altaffilmark{1}, Pei-Bin Chen\altaffilmark{1,2}, Zi-Jian Li\altaffilmark{3,4},
Cheng Cheng\altaffilmark{3,4}, Venu M. Kalari\altaffilmark{5}, Meng-Ting Shen\altaffilmark{1}, Chun-Yi Zhang\altaffilmark{1},
Junfeng Wang\altaffilmark{1}, Gaspar Galaz\altaffilmark{6}, Hong Wu\altaffilmark{3}, Zi-Qi Chen\altaffilmark{1}}

\altaffiltext{1}{Department of Astronomy, Xiamen University, 422 Siming South Road, Xiamen 361005, People$'$s Republic of China; \\astrocao@xmu.edu.cn, jfwang@xmu.edu.cn}
\altaffiltext{2}{School of Physics and Astronomy, China West Normal University, Nanchong 637009, People$'$s Republic of China}
\altaffiltext{3}{National Astronomical Observatories, Chinese Academy of Sciences, Beijing 100101, People$'$s Republic of China}
\altaffiltext{4}{Chinese Academy of Sciences South America Center for Astronomy, National Astronomical Observatories, Chinese Academy of Sciences, Beijing 100101, People$'$s Republic of China}
\altaffiltext{5}{Gemini Observatory/NSFs NOIRLab, Casilla 603, La Serena, Chile}
\altaffiltext{6}{Instituto de Astrofisica, Pontificia Universidad  Cat\'olica de Chile,\,$\!$\,Av.$\!$Vicu\~na Mackenna$\!$\,4860,\,7820436 Macul,$\!$\,Santiago, Chile}
\date{02/10/2025}

\begin{abstract}
We present integral field spectroscopy of ionized gas components
in AGC 111629, an edge-on low surface brightness galaxy (LSBG) with a stellar mass of 5.7$\times$10$^{8}$ M$_{\odot}$.
AGC 111629 displays an irregular H$\alpha$ morphology and 
an arch-like structure in the extraplanar region, which is absent in continuous stellar image.
The irregular H$\alpha$ morphology may be related to a past merger event with its satellite galaxy AGC 748815.
A peanut-shaped structure at the center in the integrated [{\ion{O}{III}}]$\lambda$5007 map,
with a position angle that differs from that of the main stellar disk.
This structure exhibits a higher [{\ion{O}{III}}]$\lambda$5007/H$\beta$ flux ratio, 
a larger equivalent width (EW) of [{\ion{O}{III}}]$\lambda$5007, and a lower H$\alpha$/H$\beta$ flux radio ($<$ 2.86). 
Some spaxels associated with the peanut-shaped structure fall within the composite region of the BPT diagram based on [{\ion{N}{II}}]$\lambda$6583. 
These features may be associated with the central AGN.
Additionally, a sub-peak in the southern disk is clearly visible in the [\ion{O}{III}]$\lambda$5007 map.
An extended region ($\sim$ 2 kpc) with an extremely low value of H$\alpha$/H$\beta$ flux ratio is observed near this sub-peak.
We interpret the sub-peak as a superbubble likely driven by supernova explosions in the southern disk.
We derive the gas-phase metallicity, 12+log(O/H), using the [\ion{N}{II}]$\lambda$6583/H$\alpha$ diagnostic and find that AGC 111629 
exhibits low central metallicity. 
This may result from feedback associated with AGN activity and supernova explosions. 

\end{abstract}

\keywords{Galaxy mergers (608); Low surface brightness galaxies (940); AGN host galaxies(2017); Supernova remnants(1667); Metallicity(1031)}

\section{Introduction \label{intro}}
Low surface brightness galaxies (LSBGs, \citealt{1987AJ.....94...23B, 2000MNRAS.312..470B, 2018ApJ...857..104G,
2021MNRAS.502.4262J}) are characterized by diffuse, faint stellar disks and are defined as having a central surface brightness
($\mu_{0}$) fainter than 23.0 mag arcsec$^{-2}$ in the $B$ band \citep{1997ARA&A..35..267I}.
These systems are dominated by dark matter (\citealt{1995MNRAS.273L..35Z, 2001AJ....122.2396D, 2002A&A...385..816D}), 
yet exhibit remarkably inefficient star formation (\citealt{2009ApJ...696.1834W, 2018ApJS..235...18L, 2024Univ...10..432C}) 
despite often harboring large reservoirs of neutral hydrogen (HI, \citealt{2015AJ....149..199D, 2024A&A...690A..69M}). 
LSBGs are typically metal-poor (\citealt{2004MNRAS.355..887K, 2017ApJ...837..152D, 2023ApJ...948...96C}) 
and deficient in molecular gas (\citealt{2001ApJ...549L.191M, 2017AJ....154..116C, 2020MNRAS.499L..26W, 2024ApJ...975L..26G}).
The fraction of active galactic nuclei (AGN) in LSBGs is lower than that in high
surface brightness galaxies (\citealt{2011ApJ...728...74G, 2020ApJ...898..106H}).
\cite{2016MNRAS.455.3148S} found LSBGs tend to have low-mass black holes (BH) and below 
the existing BH mass-velocity dispersion (M$_{\rm BH}$-$\sigma_e$) correlations.

Integral-field spectroscopy (IFS; \citealt{ALLINGTONSMITH2006244, 2008SPIE.7018E..2NV})
can simultaneously obtain spectral and spatial information on observed objects.
Taking advantage of IFS data, we can study the distribution and dynamical properties of 
ionized gas and stellar components. 
Looking into the faintEst WIth MUSE (LEWIS) is an ESO large program that conducts 
an integral-field spectroscopic survey of 30 extremely LSBGs 
in the Hydra I cluster of galaxies using MUSE at ESO Very Large Telescope (VLT)
\citep{2023A&A...679A..69I, 2025A&A...694A.276B, 2025A&A...695A..91H}.
\cite{2025arXiv250200117S} presents a spectroscopic survey of 44 targets selected
from the Systematically Measuring Ultra-Diffuse Galaxies (SMUDGes) program 
\citep{2019ApJS..240....1Z, 2021ApJS..257...60Z, 2022ApJS..261...11Z, 
2023ApJS..267...27Z, 2025OJAp....8E..90Z}, using the Keck Cosmic Web Imager (KCWI).
Additionally, there are some individual case studies 
(e.g., AGC 242019, \citealt{2021ApJ...909...20S}; Malin 1, \citealt{2024A&A...686A.247J}; 
Disco Ball, \citealt{2025ApJ...989..154K}).
These IFS observations help determine the redshifts of LSBGs and ultra-diffuse galaxies (UDGs), 
reveal environmental properties in greater detail,
and distinguish between competing formation scenarios.

However, IFS data for edge-on low surface brightness galaxies (ELSBGs) are especially scarce.
Such observations are particularly effective for detecting faint structures in galaxy outskirts, 
analyzing vertical disk features, and investigating dynamical properties.
To this end, we observed two ELSBGs (AGC 102004 and AGC 111629) 
from the sample presented by \citet{2023ApJ...948...96C}, 
using the Palomar Cosmic Web Imager (PCWI) on the 200-inch Hale Telescope at Palomar Observatory (P200). 
PCWI offers a field of view of approximately one arcminute. It employs multiple gratings 
that enable coverage of key emission lines in low-redshift targets, making it well-suited for studying extended LSBGs.
\citet{2024ApJ...971..181C} presented IFS observations of ionized gas—H$\alpha$ and [\ion{N}{II}]$\lambda$6583—in AGC 102004 
using the PCWI Red grating. From IFS data, \citet{2024ApJ...971..181C} analyzed the metallicity gradient and 
identified a potential minor or mini-merger event affecting the galaxy$'$s northwestern disk.

This paper is the second in a series presenting the analysis of IFS observations of
individual ELSBG (AGC 111629) from \cite{2023ApJ...948...96C}.
AGC 111629 is located at a redshift of 0.022 and 
has both an SDSS fiber spectrum and an HI spectrum from the ALFALFA survey.
Its major axis extends 28 kpc at a surface brightness level of 25 mag arcsec$^{-2}$ \citep{2023ApJS..269....3M}. 
AGC 111629 is detected over a broad wavelength range, spanning from the ultraviolet to the mid-infrared, 
including the far-ultraviolet (FUV), near-ultraviolet (NUV), optical ($u,g,r,i,z$), and mid-infrared (W1, W2, W3, W4) bands.
The basic parameters derived from these archival data are summarized in Table \ref{table:agc}.

In this paper, we present high-resolution ($\sim$1 kpc $\times$ 0.5 kpc) 
maps of various emission line properties (H$\beta$, [\ion{O}{III}]$\lambda$5007, H$\alpha$, and [\ion{N}{II}]$\lambda$6583) 
for AGC 111629 based on PCWI observations. 
The integral field spectroscopy observations and data reduction procedures are described in Section 2. 
Results and discussions are presented in Sections\,3 and\,4, respectively. 
Section\,5 summarizes the conclusions of this study.

\begin{table*}[h!tb]
  \caption{Basic Parameters of AGC 111629} 
  \begin{tabular}{c c c c c c c c c c c} 
  \hline\hline
  & 1  & 2 & 3 & 4 & 5 & 6 & 7& 8& 9 & 10\\ 
  \hline
  Source & Redshift & Distance  & M$_*$  & M$_{HI}$  & W50  & 12+log(O/H) &  Diameter & Inclination & b/a & Scale Length \\
  &   &  &  &  &  &  & & &  & \\
        &  &Mpc &10$^{8}$ M$_{\odot}$  & 10$^{9}$ M$_{\odot}$  &   km s$^{-1}$   &    & arcminute  & degree  &    & kpc  \\
  \hline  
  &   &  &  &  &  &  & & &  & \\ 
  AGC 111629 & 0.0215& 87.8 & 5.88 & 3.31  &  192  &   8.44    &    1.102    &   82   &  0.21  &  2.76\\
             &       & $\pm$2.1 &   & $\pm$ 0.55 & $\pm$ 3 & $\pm$ 0.007  & & &  & \\ 
  \hline
  \end{tabular}
  \begin{tablenotes} 
       \footnotesize
       \item{Col 1: The redshift is from the Siena Galaxy Atlas 2020 (SGA-2020, \citealt{2023ApJS..269....3M});}
       \item{Col 2: The distance is from ALFALFA \citep{2018ApJ...861...49H};}
       \item{Col 3: The stellar mass is from the MPA-JHU catalog \citep{2003MNRAS.341...33K};}
       \item{Col 4: The HI gas mass is form ALFALFA \citep{2018ApJ...861...49H};} 
       \item{Col 5: The velocity width of the HI line profile measured at the 50\% level of
       each of the two peaks is from ALFALFA \citep{2018ApJ...861...49H};}
       \item{Col 6: The gas phase metallicity is derived by [{\ion{N}{II}}]$\lambda$6583/H$\alpha$ diagnostic from SDSS 3$''$ fiber spectrum \citep{2023ApJ...948...96C};}
       \item{Col 7: The diameter at the 25 mag arcsec$^{-2}$ surface brightness isophote (in optical) is from SGA-2020 \citep{2023ApJS..269....3M};}
       \item{Col 8: The inclination is from \cite{2011ApJS..196...11S};}
       \item{Col 9: The semi-minor to semi-major axis ratio is from SGA-2020 \citep{2023ApJS..269....3M};}
       \item{Col 10: The scale length is measured from SDSS g-band fitted by GALFIT edge-on disk profile \citep{2023ApJ...948...96C};}
 \end{tablenotes}
  \label{table:agc} 
\end{table*}

\section{The IFS Observations and Data Reduction}

Two nights of IFS observations were conducted with PCWI on 2023 October 8 and 2024 October 
6 under the projects CTAP2023-A0053 and CTAP2024-B0041 (PI: T. Cao).
PCWI has a spectral resolution of $R = \lambda/\Delta\lambda = 5000$ 
and a field of view (FOV) of 60$''$ $\times$ 40$''$.
The instrument's spatial resolution is seeing-limited ($\sim$1.0$''$) 
along the slices in the short direction and	slit limited ($\sim$2.5$''$) in	the long direction.

We used the Red grating (640-770nm), r$'$/Red filter, and a 45nm unmasked spectral bandpass 
during the 2023 October 8 observations. 
The center wavelength of the Red grating was set to 6680 $\rm \AA$, 
corresponding to the redshifted H$\alpha$ emission.
The total on-source integration time was 4.4 hours, with a seeing of 1.1$''$.

On the other night, we used the Blue grating (460-550 nm), G$'$/Blue filter, and a 45 nm unmasked spectral bandpass.
The central wavelength of the Blue grating was set to 5054~$\rm \AA$, 
corresponding to the redshifted [{\ion{O}{III}}]$\lambda$5007 emission.
The total on-source integration time was 5.2 hours, with a seeing of 1.3$''$.

On both nights, AGC 111629 was positioned along the diagonal of the field of view (FOV) 
to maximize the coverage of the galaxy.
Each science exposure had an integration time of 20 minutes.
A blank sky background was observed off-target every 1-2 hours, 
and a standard star was also observed for calibration.

We employed the standard CWI pipeline \citep{2014ApJ...786..106M} for data reduction 
and used CWITools \citep{2020arXiv201105444O} 
to correct the World Coordinate System (WCS), coadd the WCS-corrected data, and subtract residual background emission.
The final spectral cubes have channel widths corresponding to 10.05 km s$^{-1}$ and 11.38 km s$^{-1}$ for the two nights, respectively.
The mean channel noise ($\sigma_{ch}$) is 6.0 $\times$ 10$^{-19}$ and 
5.4 $\times$ 10$^{-19}$ erg s$^{-1}$ cm$^{-2}$ $\rm \AA^{-1}$ for the two nights, respectively. 
The details of the observations are listed in Table \,\ref{table:obs}.
In the following analysis, we have corrected the observed velocity frames to the heliocentric reference frame by 
applying corrections of +5.0 km s$^{-1}$ and –1.5 km s$^{-1}$ for the two nights, respectively.

\begin{table*}[h!tb]
  \caption{Observations of AGC 111629} 
  \begin{tabular}{c c c c c c c c } 
    \hline\hline
    & & & & & & & \\
     Telescope & Instrument & Grating & Observation date  & Total on-source time & $\sigma_{ch}$ &  Seeing & Emission line  \\
     & & & & & & & \\
     & & & & hours & erg s$^{-1}$ cm$^{-2}$ $\rm \AA^{-1}$&  &\\
     \hline 
     & & & & & & & \\
       & & Red & 2023-10-08 &  4.4  &  6.0 $\times$ 10$^{-19}$ & 1.1$''$ & H$\alpha$; [{\ion{N}{II}}]$\lambda$6583 \\
      P200 & PCWI & & & & & & \\  
       &  & Blue & 2024-10-06 &  5.2  &  5.4 $\times$ 10$^{-19}$ &  1.3$''$ & H$\beta$; [{\ion{O}{III}}]$\lambda$5007,$\lambda$4959 \\
      & & & & & & & \\

  \hline
  \end{tabular}
  \label{table:obs} 
\end{table*}

\section{Results}

\subsection{H$\alpha$ and dynamical mass}
\begin{figure*}
  \begin{center}
  \includegraphics[width=7.2in]{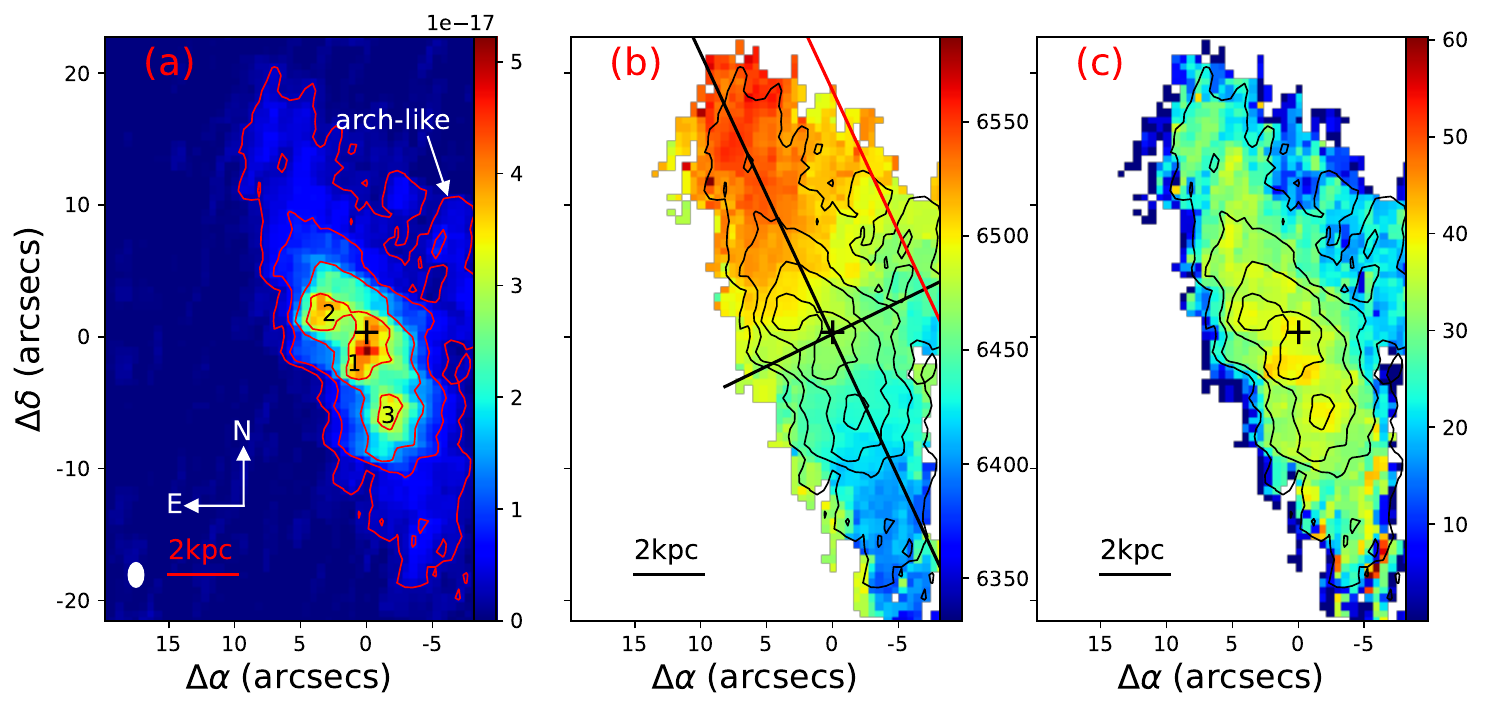}
  \end{center}
  \caption{Integrated H$\alpha$ contours at [3, 8, 18, 36] $\times$ $\sigma$ (where $\sigma$ = 8.3 $\times$ 10$^{-19}$ erg s$^{-1}$ cm$^{-2}$) 
overlaid on the images of the same integrated H$\alpha$ line emission image
in (a), the H$\alpha$ line emission velocity field (moment 1) in (b), and the velocity dispersion (moment 2) in (c). 
The images in (b) and (c) are generated by those spaxels above 3$\sigma_{ch}$, where $\sigma_{ch}$ 
is the mean channel noise ($\sigma_{ch}$ = 6.0 $\times$ 10$^{-19}$ erg s$^{-1}$ cm$^{-2}$ $\rm \AA^{-1}$) of AGC 111629. 
Three peaks of H$\alpha$ are marked by ``1", ``2", and ``3".
We mark the major axis and minor axis with black lines in panel (b) with 25 degree PA. 
The red line parallels to major axis and at 8.0$''$ offset from the perpendicular disk center.
The black cross represents the center of integrated H$\alpha$ image fitted by GALFIT 
S$\acute{e}$rsic profile. The oval in the lower left corner represents the spatial resolution 2.5$''$ $\times$ 1.1$''$.
}
  \label{Ha}
\end{figure*}

Panels (a), (b), and (c) in Figure\,\ref{Ha} depict the integrated H$\alpha$ image,
the velocity field, and the velocity dispersion maps (i.e., moment 0, moment 1, and moment 2 \href{https://www.aoc.nrao.edu/~kgolap/casa_trunk_docs/CasaRef/image.moments.html}{$^1$})
 of AGC 111629, respectively.
We apply $^{3D}$BAROLO \citep{2015MNRAS.451.3021D} to obtain the velocity field and the velocity dispersion maps.
The center of the integrated H$\alpha$ image (indicated by the plus marker in Figure\,\ref{Ha}) 
is determined from the GALFIT \citep{2010AJ....139.2097P} S$\acute{e}$rsic profile fitting. 
The position angle (PA) derived from the S$\acute{e}$rsic profile is 25 degrees (measured North-to-East) for the integrated H$\alpha$ image, 
corresponding to the major axis direction shown in Figure\,\ref{Ha}(b).
Three peaks (labeled ``1", ``2", and ``3" in Figure\,\ref{Ha}(a)) are identified in the integrated H$\alpha$ image at 36$\sigma_{H\alpha}$ profile, 
with an arch-like structure extending along the H$\alpha$ minor axis at 3$\sigma_{H\alpha}$ profile in the extraplanar region.
The arch-like structure exhibits a velocity distribution similar to that of the disk plane but with lower velocity dispersion.
Compared to another ELSBG AGC 102004 \citep{2024ApJ...971..181C}, 
the morphology of the integrated H$\alpha$ image of AGC 111629 appears more irregular.

\begin{figure}
  \begin{center}
  \includegraphics[width=3.5in]{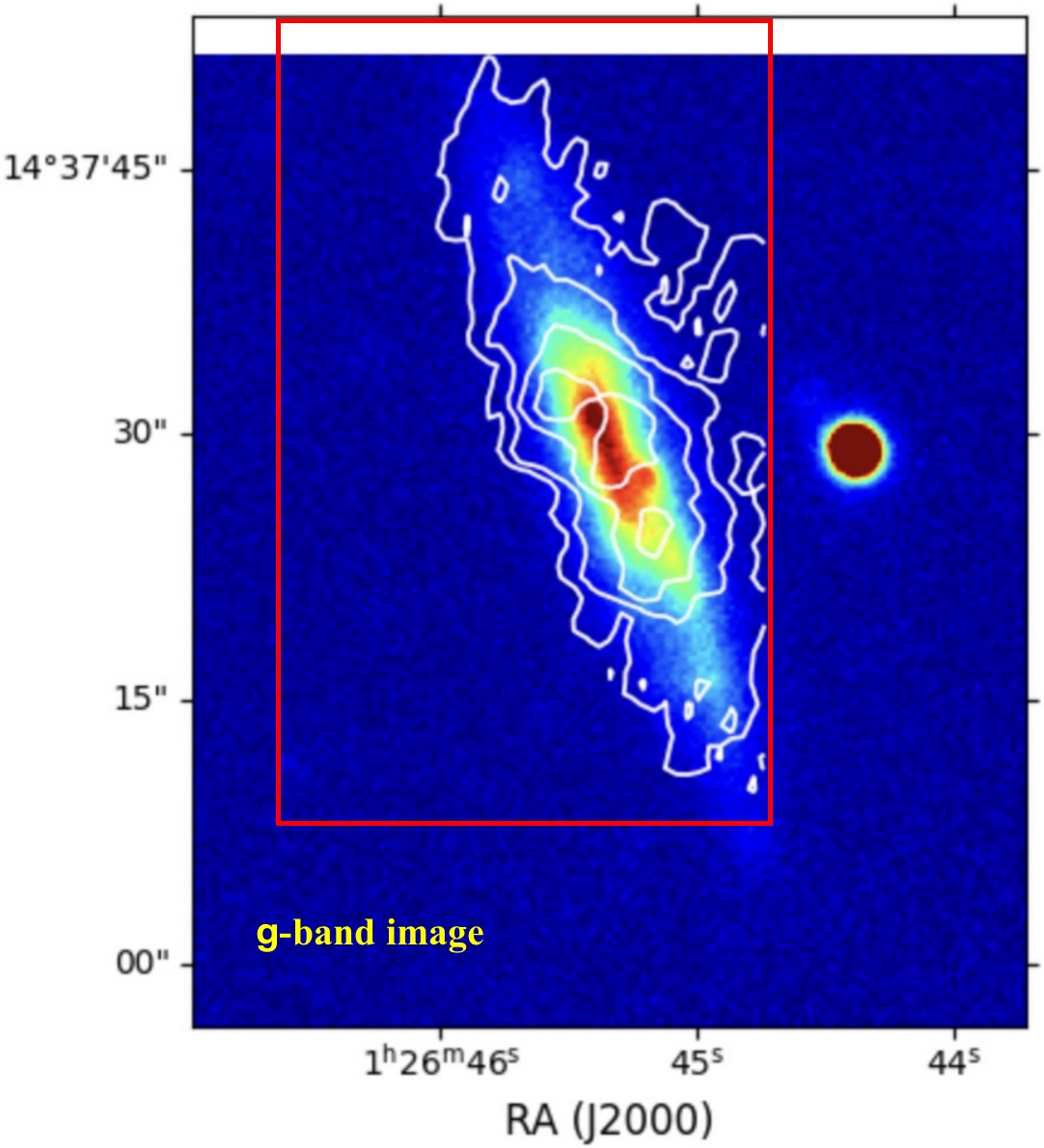}
  \end{center}
  \caption{Integrated H$\alpha$ emission contours (white lines) at [3, 8, 18, 36] $\times$ $\sigma$ 
  (where $\sigma$ = 8.3 $\times$ 10$^{-19}$ erg s$^{-1}$ cm$^{-2}$) overlaid on DESI g-band image of AGC 111629.
  The point source near AGC 111629 is identified as a star in SDSS. 
  The red square is the same region of Figure \ref{Ha}.
  }
  \label{gband}
\end{figure}

\begin{figure}
  \begin{center}
  \includegraphics[width=3.5in]{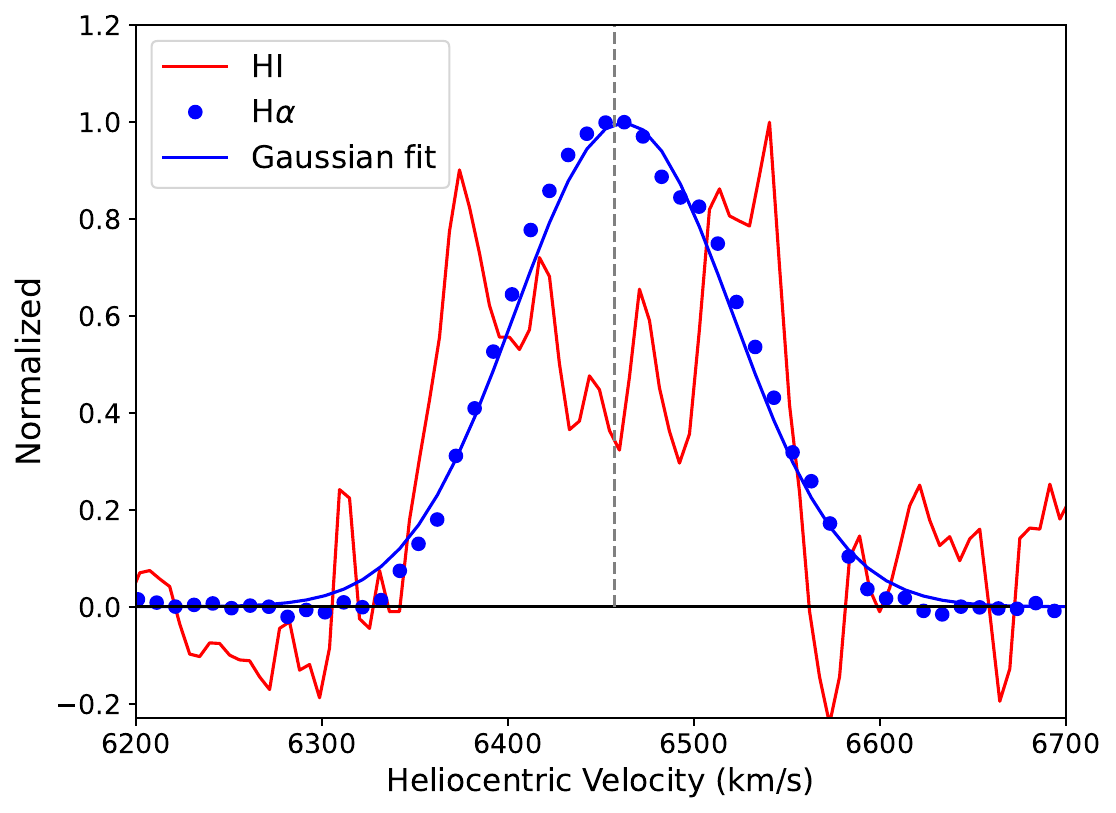}
  \end{center}
  \caption{The red line is the HI spectrum from ALFALFA, and the blue dots represent
  the integrated H$\alpha$ spectrum of AGC 111629 among the 3$\sigma$ contour region in Figure\,\ref{Ha}(a).
  The blue solid line is the Gaussian fitting of the integrated H$\alpha$ spectrum.
  The grey dashed line is the systemic velocity of the Gaussian fitting center 6457 km s$^{-1}$ 
  which is consistent with the center of HI spectrum 6454 km s$^{-1}$}.
  \label{spectra}
\end{figure}

The overall distribution of the H$\alpha$ emission closely matches the continuous stellar image, 
except for the arch-like structure, which is not visible in the $g$-band (see Figure\,\ref{gband}).
Figure\,\ref{spectra} shows the H$\alpha$ spectrum of AGC 111629 within the 3$\sigma$ contour region in Figure\,\ref{Ha}(a).
A single Gaussian component provides a good fit to the spectrum, yielding a central velocity of 6457 km s$^{-1}$ 
and a velocity dispersion ($\sigma$) of 51.64 km s$^{-1}$.
The H$\alpha$ emission is consistent with the HI spectrum from ALFALFA.

\begin{figure*}
  \begin{center}
  \includegraphics[width=7.0in]{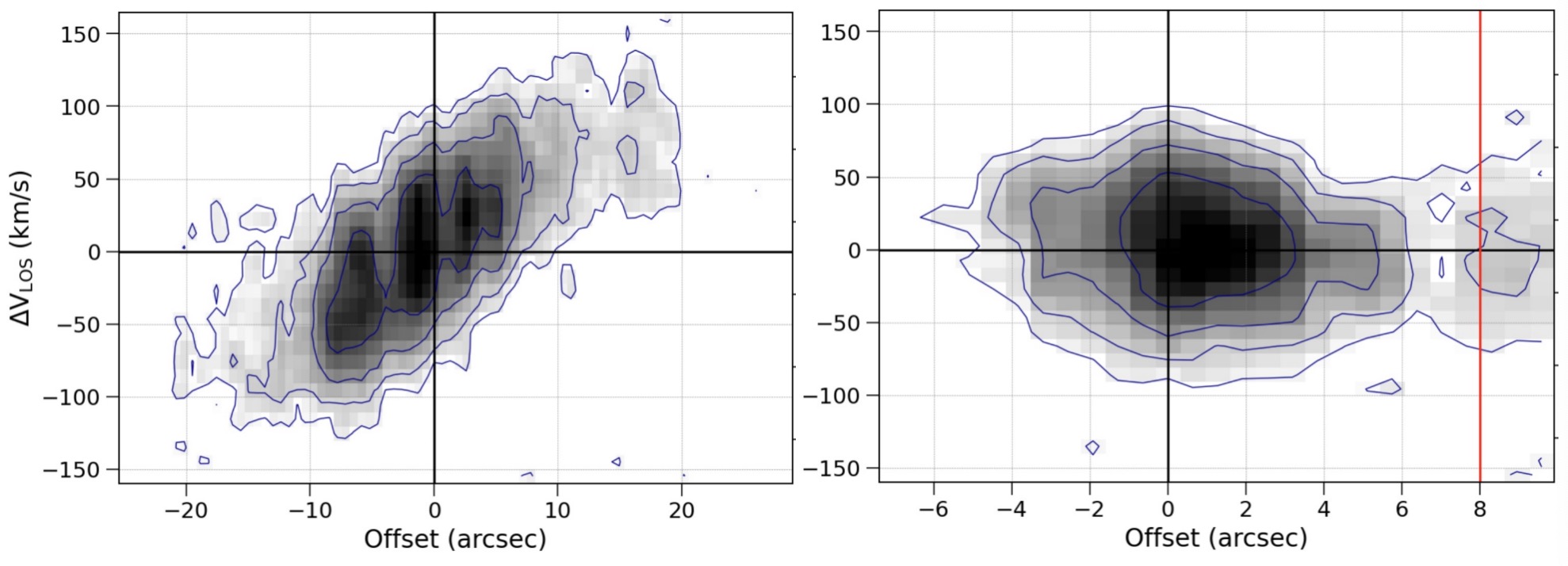}
  \end{center}
  \caption{Left panel:The H$\alpha$ PVD along the major axis of AGC 111629. H$\alpha$ is shown in
  gray with blue contours. Right panel: The H$\alpha$ PVD along the minor axis of AGC 111629. 
  The red line marks the arch-like structure position in Figure\,\ref{Ha}(b). 
  1$''$ corresponds to 0.43 kpc.
  }
  \label{PVD}
\end{figure*}

We utilize $^{3D}$BAROLO to generate the position-velocity diagram (PVD).
Overall, the H$\alpha$ emission exhibits characteristics of a rotating disk in the PVD along the major axis (PA = 25 degrees) 
as shown in the left panel of Figure\,\ref{PVD}, although the structure is highly asymmetric.
The arch-like structure is clearly visible in the PVD along the minor axis in the right panel of Figure\,\ref{PVD}, 
marked by a red line located at an offset of 8$''$ ($\sim$ 3.4 kpc) from the center along the perpendicular direction to the disk.

The dynamical mass (M$\rm_{dyn}$) of galaxy can be estimated from the rotation velocity (V$_{\rm rot}$) in
combination with an assumed or measured size of the HI disk (R$_{\rm HI}$) (e.g. \citealt{1980A&A....81..371C, 2008ApJS..177..103H, 2018MNRAS.478.1611K}). 
The dynamical mass inferred from HI rotation is given by \cite{2020ApJ...898..102Y}:
\begin{equation}
  \rm M_{dyn} = \frac{V^2_{rot}R_{HI}}{G} = 2.31 \times 10^5\,M_{\odot}\,\frac{V^2_{rot}}{km s^{-1}}\,\frac{R_{HI}}{kpc}
\end{equation}
The rotation velocity is measured from the width of the HI line profile at 
50\% of the peak flux (W50), corrected for inclination as V$_{\rm rot}$ = W50/2sin(i)
\citep{2019MNRAS.483.1754D}. Since single-dish spectra do not provide direct measurements of R$_{\rm HI}$,
we adopt the empirical relation between R$_{\rm HI}$ and HI mass from \cite{2016MNRAS.460.2143W}
 to estimate R$_{\rm HI}$ for AGC 111629.

M$\rm_{dyn}$ of AGC 111629 is 4.04$\times$10$^{10}$ M$_{\odot}$, which is 
an order of magnitude higher than its HI mass. 
The ratio of stellar mass to dynamical mass is 0.014, indicating that AGC 111629 is strongly dark matter dominated.

\subsection{[{\ion{O}{III}}]$\lambda$5007 and H$\beta$}

\begin{figure*}
  \begin{center}
  \includegraphics[width=7.2in]{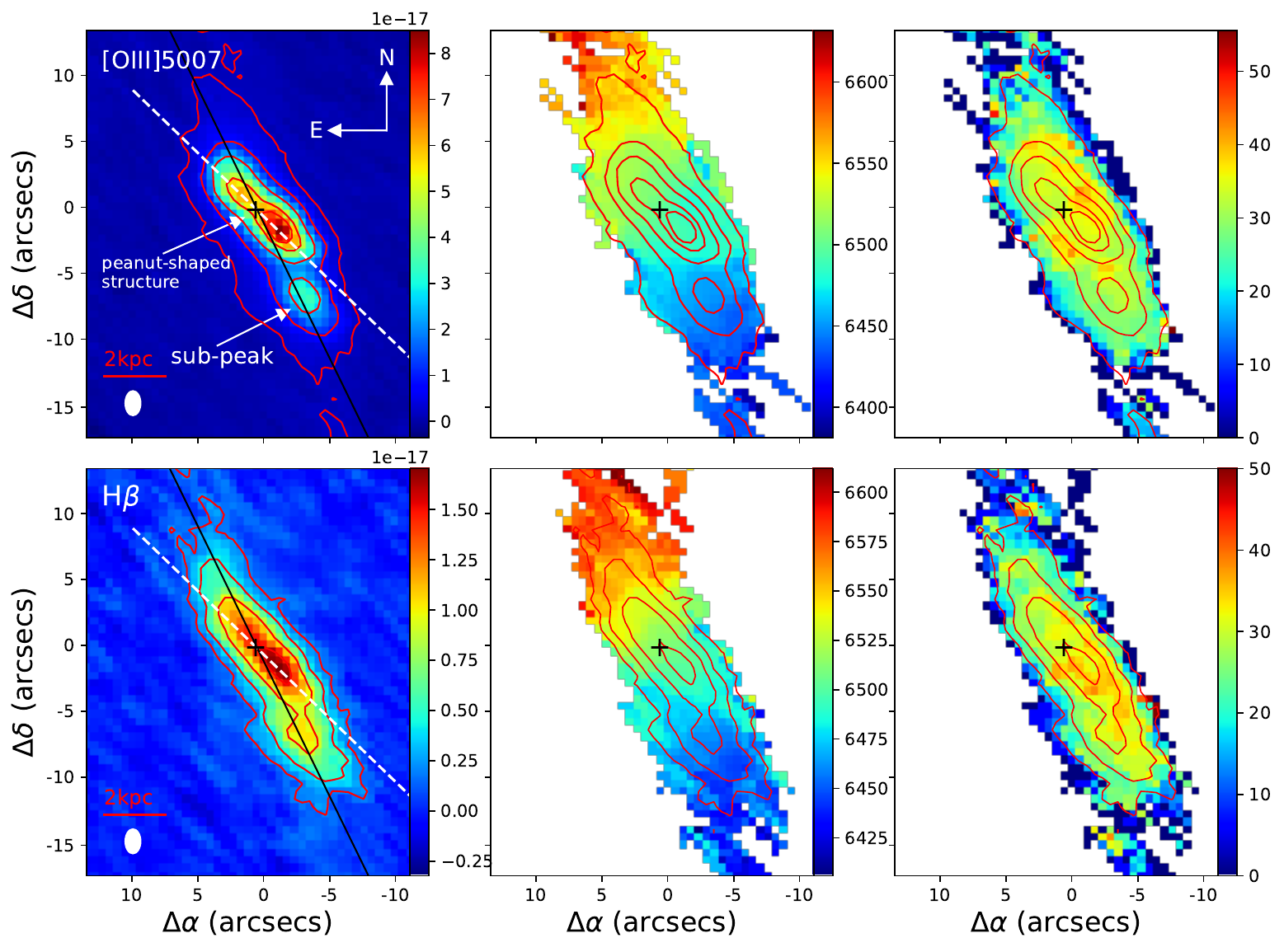}
  \end{center}
  \caption{The upper panels: Integrated [{\ion{O}{III}}]$\lambda$5007 contours at [3, 15, 40, 80, 105] 
  $\times$ $\sigma$ (where $\sigma$ = 6.2 $\times$ 10$^{-19}$ erg s$^{-1}$ cm$^{-2}$) 
overlaid on the images of the same integrated [{\ion{O}{III}}]$\lambda$5007 line emission image
in the left, the [{\ion{O}{III}}]$\lambda$5007 emission line velocity field (moment 1) in the middle, and the velocity dispersion (moment 2) in the right. 
The lower panels: Integrated H$\beta$ contours at [3, 7, 15, 25] $\times$ $\sigma$ 
(where $\sigma$ = 6.0 $\times$ 10$^{-19}$ erg s$^{-1}$ cm$^{-2}$) 
overlaid on the images of the same integrated H$\beta$ line emission image
in the left, the H$\beta$ emission line velocity field (moment 1) in the middle, and the velocity dispersion (moment 2) in the right.
The images in velocity fields (moment 1) and the velocity dispersions (moment 2) are generated by 
those spaxels above 3$\sigma_{ch}$, where $\sigma_{ch}$ 
is the mean channel noise ($\sigma_{ch}$ = 5.4 $\times$ 10$^{-19}$ erg s$^{-1}$ cm$^{-2}$ $\rm \AA^{-1}$). 
The black cross represents the same the center of integrated H$\alpha$ image with Figure\,\ref{Ha}.
The black line is the major axis of H$\alpha$ image with PA = 25 degree. 
The PA of the white dashed line is 46 degree.
The oval in the lower left corner represents the spatial resolution 2.5$''$ $\times$ 1.3$''$.}
  \label{Hb}
\end{figure*}

We present the integrated line emission images, velocity fields, 
and velocity dispersion maps for [{\ion{O}{III}}]$\lambda$5007 and H$\beta$ in Figure\,\ref{Hb}.
The [{\ion{O}{III}}]$\lambda$5007 emission exhibits 
a peanut-shaped structure in the central region and a sub-peak in the southern disk.
The peanut-shaped structure has a different position angle (white dashed line in Figure\,\ref{Hb}) 
compared to the entire galaxy (black line in Figure\,\ref{Hb}).
[{\ion{O}{III}}]$\lambda$5007 and H$\beta$ display similar morphologies 
around the peanut-shaped structure and the sub-peak.
The velocity dispersions around these structures are 
higher than in other regions.

\begin{figure*}
  \begin{center}
  \includegraphics[width=7.0in]{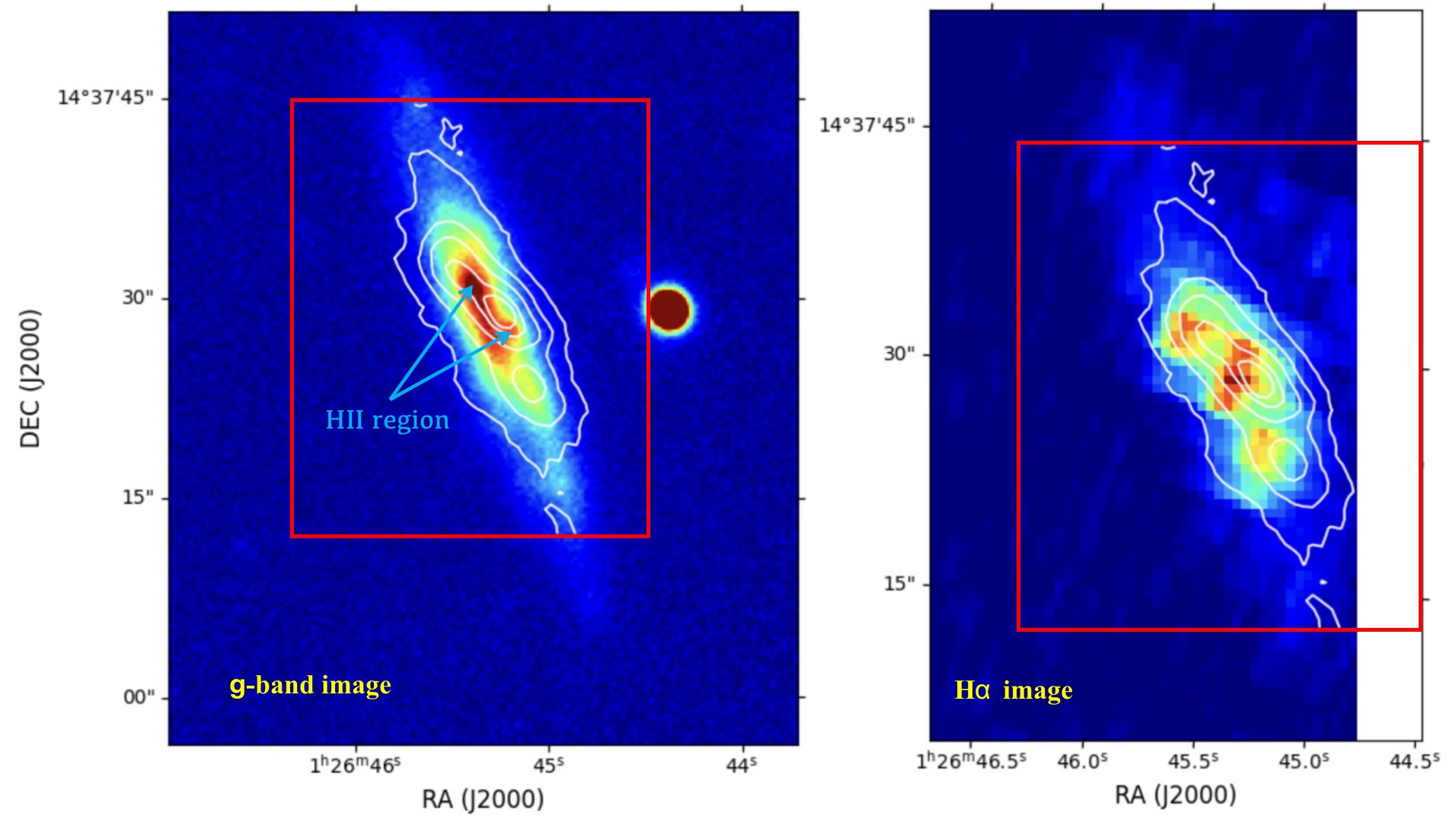}
  \end{center}
  \caption{Integrated [{\ion{O}{III}}]$\lambda$5007 emission contours (white lines) at [3, 15, 40, 80, 105] $\times$ $\sigma$ 
  (where $\sigma$ = 6.2 $\times$ 10$^{-19}$ erg s$^{-1}$ cm$^{-2}$) 
  overlaid on DESI g-band image (in the left panel) and integrated H$\alpha$ image (in the right panel) of AGC 111629.
  The red square is the same region of Figure \ref{Hb}.}
  \label{O3gband}
\end{figure*}

The morphology of the peanut-shaped structure in the central region differs 
from that of the stellar continuum shown in Figure\,\ref{O3gband} 
and near to two HII regions visible in the DECalS legacy survey image of the galaxy.
Additionally, the peanut-shaped structure and sub-peak in [{\ion{O}{III}}]$\lambda$5007
 and H$\beta$ are slightly offset from the H$\alpha$ peaks, 
as seen in Figure\,\ref{O3gband}.
No 3$\sigma$ detection of [{\ion{O}{III}}]$\lambda$5007 or H$\beta$ 
is found at the location of the H$\alpha$ arch-like structure (Figure\,\ref{Ha}(a)).
We note that the observational depth of the [{\ion{O}{III}}] and H$\alpha$ data is comparable.

\subsection{Lines ratios}

\subsubsection{BPT diagram}
\begin{figure}
  \begin{center}
  \includegraphics[width=3.5in]{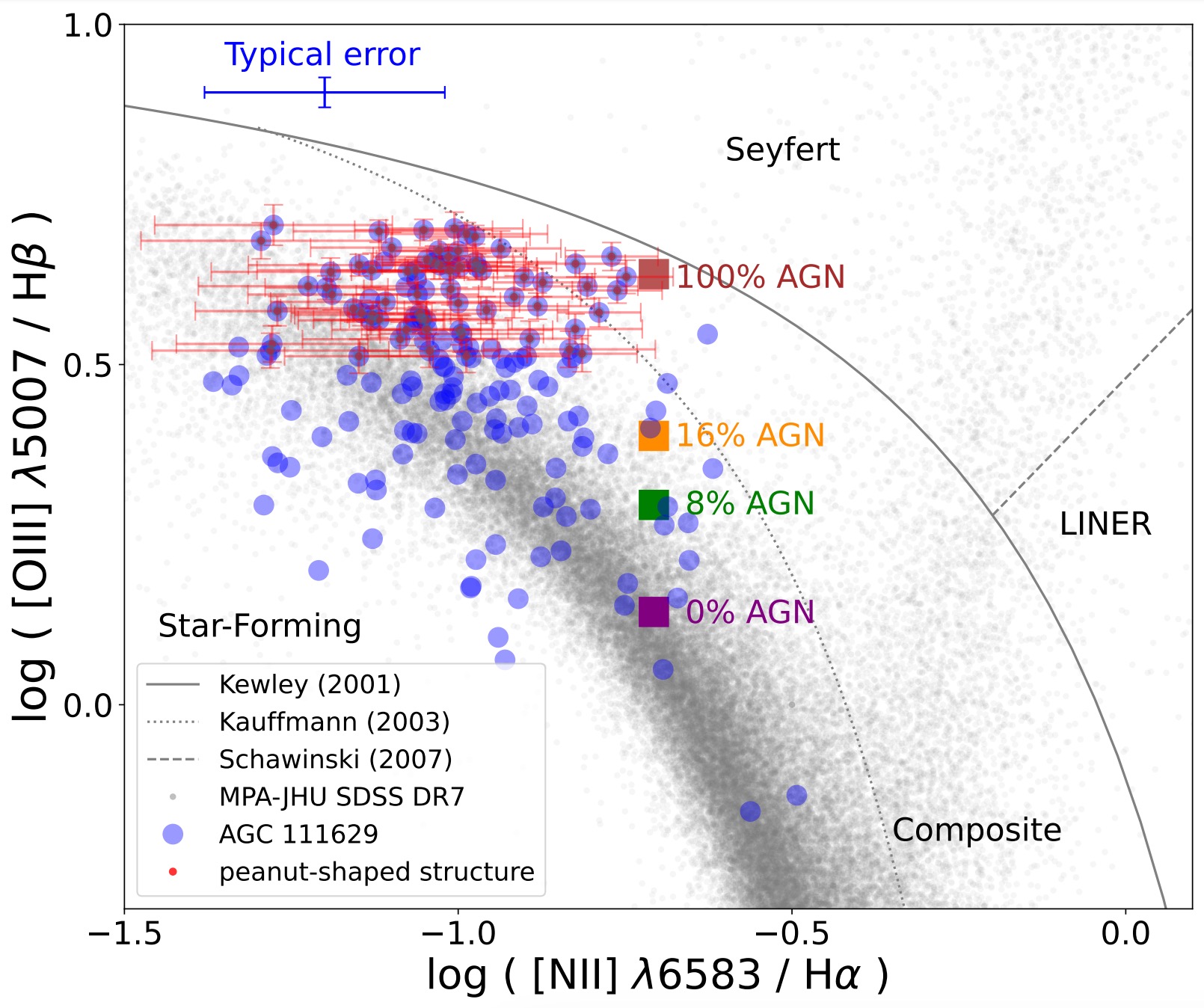}
  \end{center}
  \caption{BPT diagram.
  The blue dots represent each spaxel of AGC 111629 within 3$\sigma$ [{\ion{N}{II}}]$\lambda$6583 and H$\beta$ region 
  as show in Figure\,\ref{metal}(b). The red dots with errors represent the center peanut-shaped structure within 
  [{\ion{O}{III}}]$\lambda$5007 40$\sigma$ region (see Figure\,\ref{3b}). 
  The typical error is the mean value of all blue dots.
  Gray dots represent galaxies from the MPA-JHU SDSS DR7 \citep{2004MNRAS.351.1151B}.
  The gray solid line is from \cite{2001ApJ...556..121K}, 
  the dotted line is from \cite{2003MNRAS.346.1055K}, and the dashed line is from \cite{2007MNRAS.382.1415S}.
  The model data points (purple, green, darkorange, brown squares) are from \cite{2022ApJ...931...44P} for a low-metallicity
theoretical dwarf galaxy with a continuous star formation history (CSF), 
a metallicity of Z = 0.4Z$_{\odot}$, and a 0\%, 8\%, 16\%, or 100\% AGN contribution 
to its spectrum from a black hole with M$_{\rm BH}$ = 10$^{5.0}$ M$_{\odot}$.}
      \label{BPT}
\end{figure}

The Baldwin-Phillips-Terlevich (BPT) diagram is a
powerful tool for characterizing extragalactic properties
based on various emission-line strength ratios \citep{1981PASP...93....5B}.
We present the spatially resolved BPT diagram ([{\ion{O}{III}}]$\lambda$5007/H$\beta$ vs. [{\ion{N}{II}}]$\lambda$6583/H$\alpha$) 
of AGC 111629 within the 3$\sigma$ detection region of [{\ion{N}{II}}]$\lambda$6583 and H$\beta$ in Figure\,\ref{BPT}.
This 3$\sigma$ detection region mainly covers the three H$\alpha$ peaks (see Figure\,\ref{metal}).
Most points fall within the star-forming region.
A few points associated with the peanut-shaped structure lie in the composite region and 
show a smaller error than the typical error.
The SDSS data points represent averages over the central regions of galaxies, whereas our data cover different spatial scales, 
which may account for the offset observed relative to the SDSS points in Figure\,\ref{BPT} (e.g. \citealt{2020ARA&A..58...99S}).

We note that the BPT diagram based on [{\ion{N}{II}}]$\lambda$6583 is metallicity sensitive \citep{2022ApJ...931...44P}.
AGN in low metallicity dwarfs are likely placed in the star-forming region of the BPT diagram based on [{\ion{N}{II}}]$\lambda$6583 plot
\citep{2006MNRAS.371.1559G, 2012ApJ...756...51L}, as the example model data points shown in Figure\,\ref{BPT}.

\begin{figure}
  \begin{center}
  \includegraphics[width=3.5in]{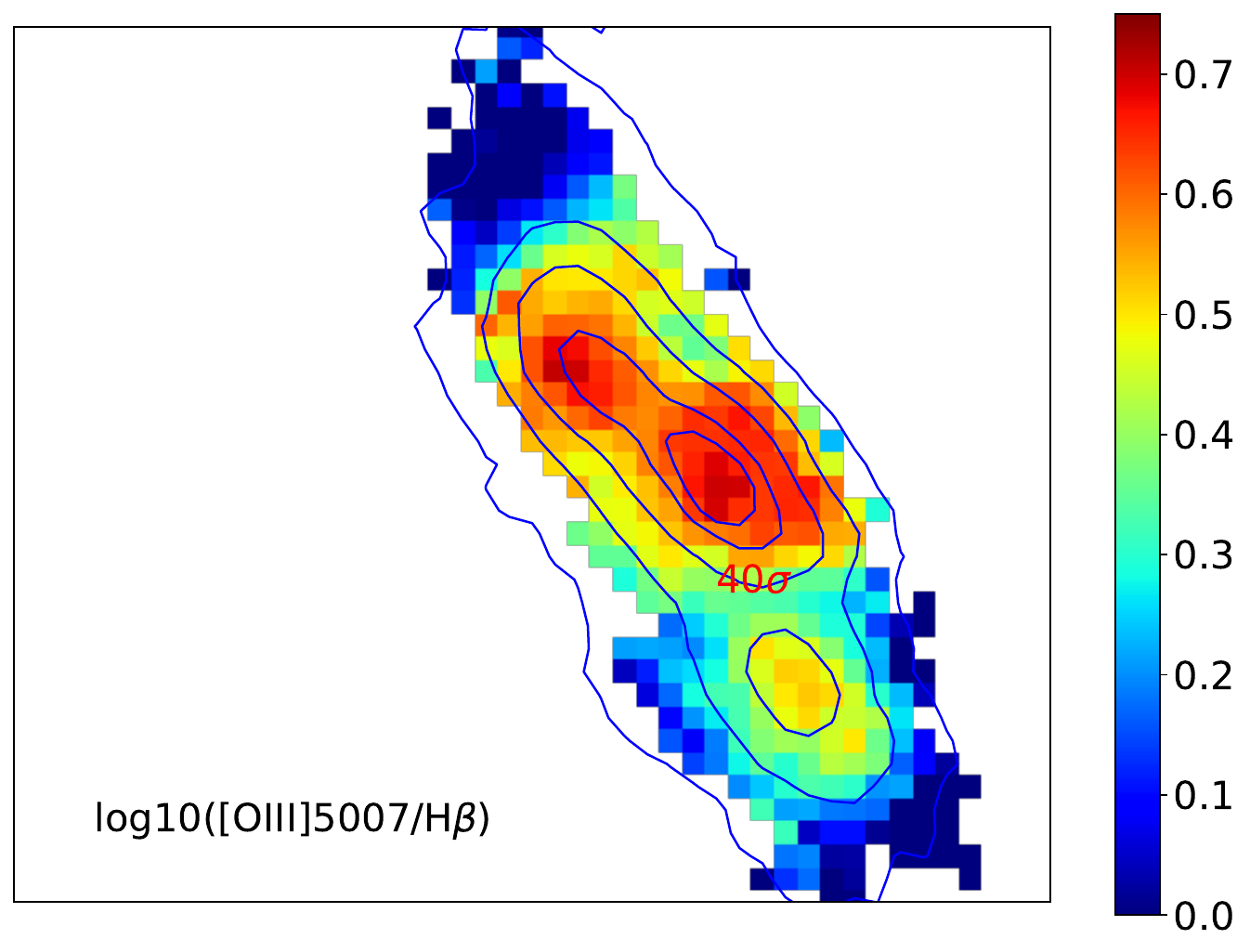}
  \end{center}
  \caption{[{\ion{O}{III}}]$\lambda$5007 to H$\beta$ flux ratio within 3$\sigma$ H$\beta$ region. 
  The blue contours (integrated [{\ion{O}{III}}]$\lambda$5007 at [3, 15, 40, 80, 105] 
  $\times$ $\sigma$, where $\sigma$ = 6.2 $\times$ 10$^{-19}$ erg s$^{-1}$ cm$^{-2}$) are the same as 
  [{\ion{O}{III}}]$\lambda$5007 contours in Figure\,\ref{Hb} and Figure\,\ref{O3gband}.}
  \label{3b}
\end{figure}

We present the [{\ion{O}{III}}]$\lambda$5007 to H$\beta$ ratio ([{\ion{O}{III}}]$\lambda$5007/H$\beta$) map in Figure\,\ref{3b}.
Within the peanut-shaped structure, the [{\ion{O}{III}}]$\lambda$5007/H$\beta$ map shows high values.
While, the sub-peak of [{\ion{O}{III}}]$\lambda$5007 exhibits lower values in the [{\ion{O}{III}}]$\lambda$5007/H$\beta$ map.
The different values of [{\ion{O}{III}}]$\lambda$5007/H$\beta$ suggests different ionization environments between the peanut-shaped structure 
and the sub-peak seen in the [{\ion{O}{III}}]$\lambda$5007 image.

\subsubsection{The Balmer Decrement}

\begin{figure*}
  \begin{center}
  \includegraphics[width=6.2in]{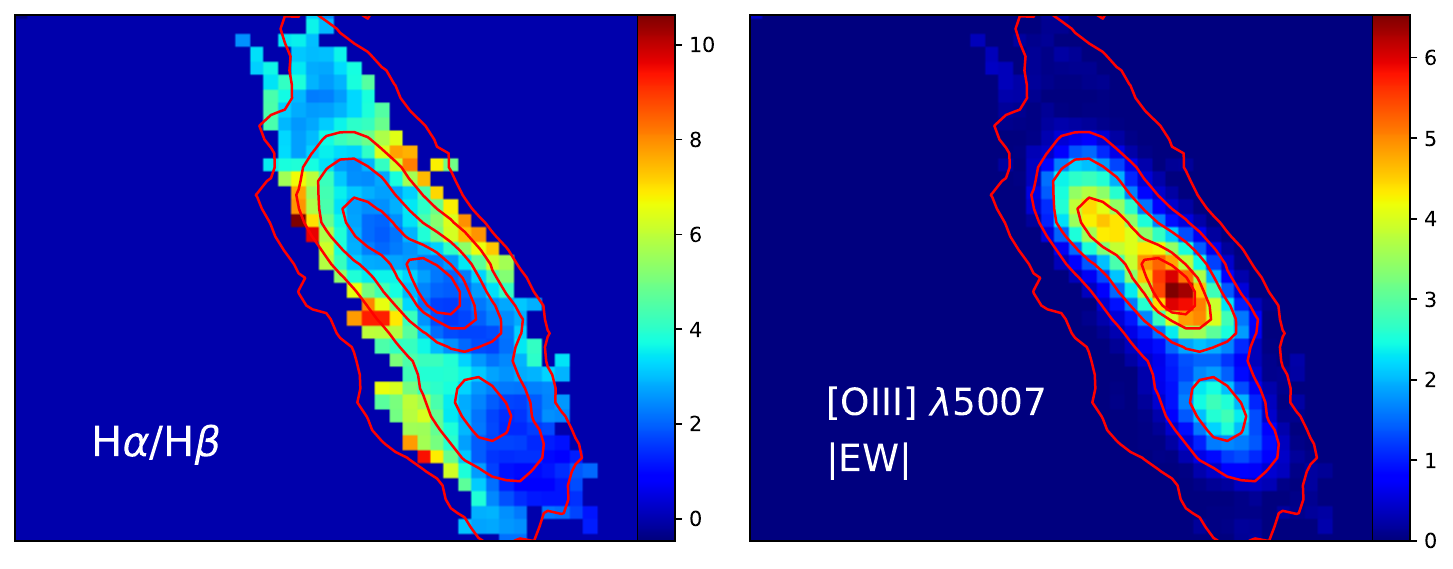}
  \end{center}
  \caption{Left panel: H$\alpha$ to H$\beta$ flux ratio within 3$\sigma$ H$\beta$ region. 
  The red contours (integrated [{\ion{O}{III}}]$\lambda$5007 at [3, 15, 40, 80, 105] 
  $\times$ $\sigma$, where $\sigma$ = 6.2 $\times$ 10$^{-19}$ erg s$^{-1}$ cm$^{-2}$) are the same as 
  [{\ion{O}{III}}]$\lambda$5007 contours in Figure\,\ref{Hb}, Figure\,\ref{O3gband}, and Figure\,\ref{3b}.
  The resolution of integrated H$\alpha$ and H$\beta$ images have been convolved to the same value (2.5$''$ $\times$ 1.5$''$).
  Right panel: The absolute value of Equivalent Width ($|\rm EW|$) of [{\ion{O}{III}}]$\lambda$5007. The figure is masked by H$\beta$ 3$\sigma$ region for 
  better comparison with H$\alpha$ to H$\beta$ flux ratio. The red contours are the same as the left panel. 
  The $|\rm EW|$ is in the unit of $\rm \AA$.}  
  \label{hab}
\end{figure*}

The Balmer decrement (i.e., H$\alpha$/H$\beta$) is commonly used to determine the attenuation in galaxies.
We present the H$\alpha$ to H$\beta$ ratio (H$\alpha$/H$\beta$) map in the left panel of Figure\,\ref{hab}.
The native resolutions of the H$\alpha$ and H$\beta$ images are 2.5$''$ $\times$ 1.1$''$ and 2.5$''$ $\times$ 1.3$''$, respectively.
To create a consistent flux ratio map, we convolved both images to a common resolution of 2.5$''$ $\times$ 1.5$''$ using a Gaussian kernel.

To account for potential continuum absorption and calibration offsets between the blue and red spectra, we compared the H$\alpha$ and H$\beta$ 
fluxes from the SDSS 3$''$ fiber spectrum with those measured from our data over the same region. 
The fluxes and flux ratios show good agreement, with both datasets yielding H$\alpha$/H$\beta \approx 2.7$. 
This consistency indicates that continuum absorption and calibration differences have only a negligible impact on our measurements.

In galaxies, extinction in the central region is typically higher than in the outer disk, resulting in a higher H$\alpha$/H$\beta$ ratio at the center.
However, for AGC 111629, we find an anomalous result in Figure\,\ref{hab}.
The central peanut-shaped structure exhibits a lower H$\alpha$/H$\beta$ value compared to the outskirts along the minor axis.
Additionally, the H$\alpha$/H$\beta$ ratios in the southern and northern disks are asymmetric.
In the northern disk, H$\alpha$/H$\beta$ is approximately 3.0, whereas in the southern disk, it is mostly below 2.5.
In AGC 111629, both the central peanut-shaped structure and the southern disk show H$\alpha$/H$\beta$ values lower than the canonical Case B value.
Although unusual, there are documented cases of H$\alpha$/H$\beta$ ratios below 2.86 in galaxies 
(e.g., \citealt{2009A&A...506L...1A, 2015ApJ...806..259R, 2017ApJ...847...38Y, 2021ApJ...923....6J, 2024A&A...681A.100J})
and some of which are as small as $\sim$1.0 \citep{2025arXiv250321896S}.

The right panel shows the absolute value of Equivalent Width ($|\rm EW|$) of [{\ion{O}{III}}]$\lambda$5007.
Both the peanut-shaped structure and sub-peak show strong [{\ion{O}{III}}]$\lambda$5007 emission (higher $|\rm EW|$ value).
The H$\alpha$/H$\beta$ flux ratio no longer satisfies the conditions for the canonical Case B value in those regions.

\subsubsection{Metallicity}

\begin{figure*}
  \begin{center}
  \includegraphics[width=7.0in]{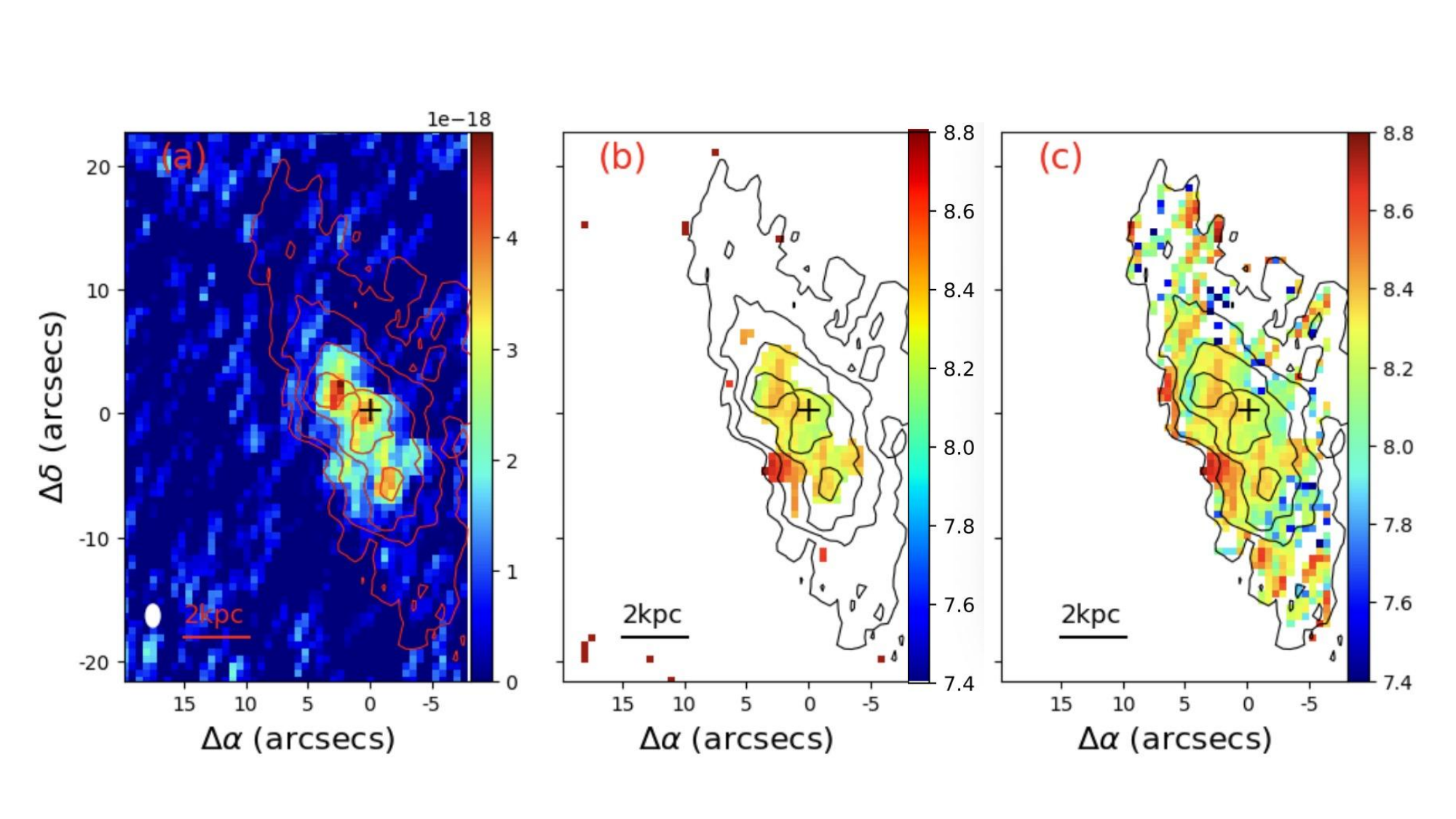}
  \end{center}
\caption{Integrated H$\alpha$ contours at [3, 8, 18, 36] $\times$ $\sigma$ 
(where $\sigma$ = 8.3 $\times$ 10$^{-19}$ erg s$^{-1}$ cm$^{-2}$) are overlaid on the
integrated [{\ion{N}{II}}] line emission image in panel (a) and on the gas-phase abundance maps of
 12+log(O/H) in panels (b) and (c).
The metallicity in panel (b) is derived using only spaxels with [{\ion{N}{II}}] detections above
3$\sigma_{[NII]}$ (1 $\sigma_{[NII]}$ = 7.0 $\times$ 10$^{-19}$ erg s$^{-1}$ cm$^{-2}$), 
while panel (c) includes spaxels with lower signal-to-noise ratios.
The oval in the lower left corner represents the spatial resolution 2.5$''$ $\times$ 1.1$''$.}
   \label{metal}
\end{figure*}

Panel (a) of Figure\,\ref{metal} shows the integrated [{\ion{N}{II}}]$\lambda$6583 emission image, 
while panel (b) presents the derived metallicity, 12 + log(O/H), 
based on the [{\ion{N}{II}}]$\lambda$6583/H$\alpha$ ratio (the [N2] index; \citealt{2004MNRAS.348L..59P}).
The three peaks in the integrated [{\ion{N}{II}}]$\lambda$6583 image 
are spatially consistent with the H$\alpha$ peaks at 36$\sigma_{H\alpha}$.
The metallicity estimation includes only spaxels with 
[{\ion{N}{II}}]$\lambda$6583 detection above 3$\sigma$ in Figure\,\ref{metal}(b).
The metallicities range from 8.18 to 8.70, with a mean value of 8.31.
The central H$\alpha$ peak ``1" exhibits a slightly lower metallicity 
compared to the regions around the other two peaks. 
The drop in metallicity of H$\alpha$ peak ``1" may be associated with enhanced star-formation 
(e.g., \citealt{2013ApJ...767...74S, 2019ApJ...882....9S}).

We further assessed the metallicity in regions with low [{\ion{N}{II}}] signal. 
In Figure\,\ref{metal}(c), the metallicities range from 7.4 to 8.8.
As shown in Figure\,\ref{metal}(c), the metallicities range from 7.4 to 8.8.
Significant [\ion{N}{II}] detections (above 3$\sigma$) are primarily concentrated in the central region, 
while the 12 + log(O/H) values in the outer disk should be considered upper limits.
Notably, the outer region on the eastern side of the vertical disk exhibits, on average, 
higher metallicity than the western side, which is connected to the arch-like structure.

\section{Discussions}

\subsection{A past merger event in AGC 111629}

Irregularities in H$\alpha$ morphology are fairly common in star-forming galaxies 
and are often observed in isolated systems (e.g., tadpole galaxies, \citealt{2016ApJ...825..145E};
 or Blue Compact Dwarf galaxies (BCDs), \citealt{2004AJ....128.2170H}, \citealt{2010A&A...520A..90C}).
It has been suggested that BCDs can form through interactions or mergers between gas-rich dwarf galaxies
\citep{2020ApJ...891L..23Z, 2023MNRAS.520.4953C, 2024ApJ...965....3Z}.

AGC 111629 is a member of a galaxy group (Group ID: 10957) which includes three galaxies in \cite{2007ApJ...671..153Y},
with a total stellar mass of $\sim10^{10}$ M$_{\odot}$ and a halo mass of $\sim10^{12}$ M$_{\odot}$.
Both the central galaxy and another member galaxy are located northeast of AGC 111629.
AGC 111629 lies at projected distances of approximately 10$'$ (240 kpc) and 15$'$ (360 kpc) 
from the central and the other member galaxy, respectively. 

\begin{figure*}
  \begin{center}
  \includegraphics[width=7.0in]{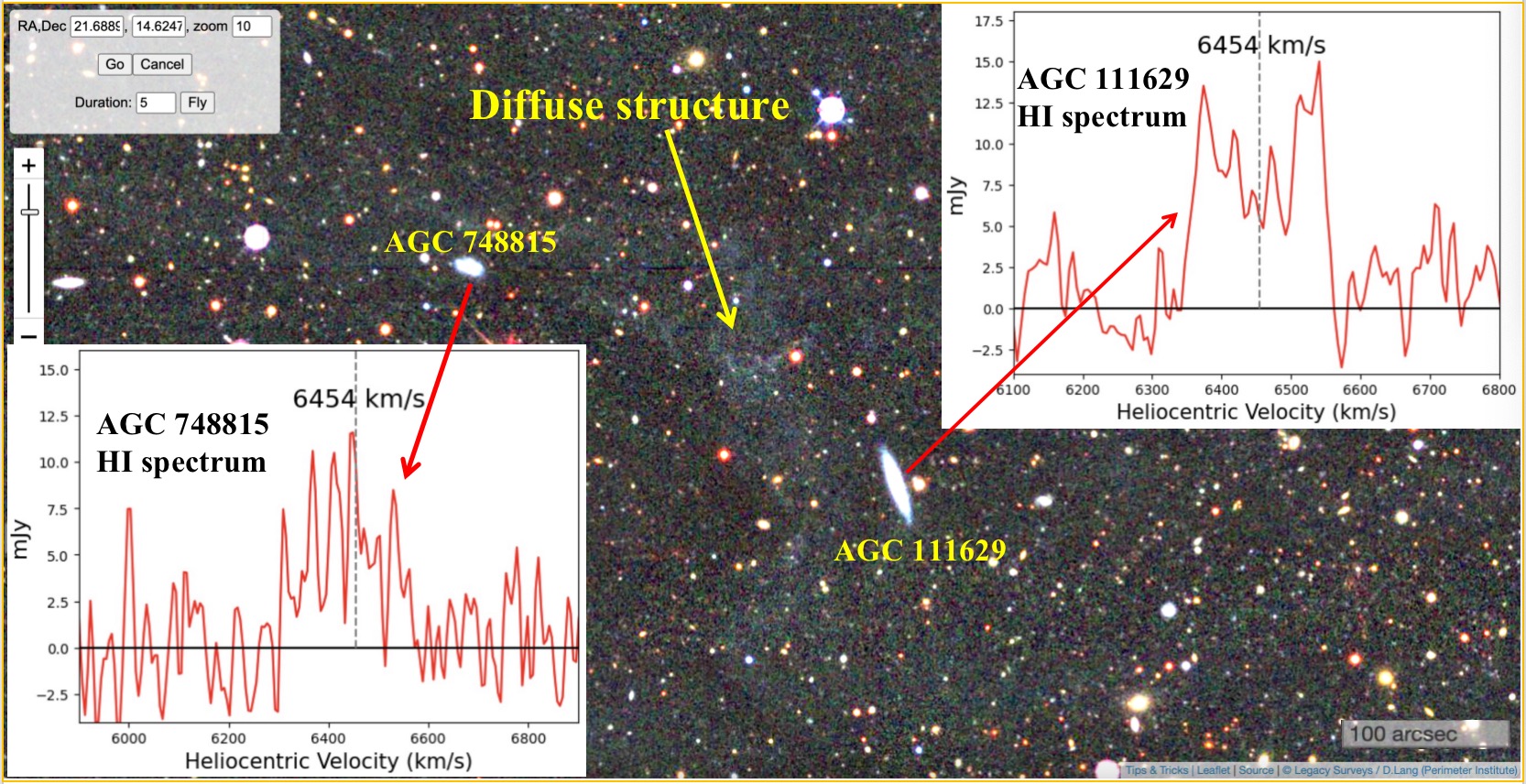}
  \end{center}
  \caption{Color image of AGC 111629 and its satellite galaxy AGC 748815, 
  based on smoothed Legacy Survey DR9 data (https://www.legacysurvey.org/viewer). 
  Both galaxies are marked. The HI spectra from ALFALFA show that the two galaxies share the same velocity. 
  A diffuse structure is clearly visible between AGC 111629 and AGC 748815. 1$''$ corresponds to 0.43 kpc.} 
  \label{sate}
\end{figure*}

Furthermore, we confirmed a satellite galaxy (AGC 748815) associated with AGC 111629 
as shown in Figure\,\ref{sate} by HI velocity. 
The HI spectra from ALFALFA of AGC 111629 and AGC 748815 indicate they are at the same redshift. 
A diffuse structure is observed between the two galaxies, which may represent the remnant of a past merger. 
We speculate that a merger between AGC 111629 and AGC 748815 contributed to the irregular H$\alpha$ morphology in AGC 111629. 
The projected separation between the galaxies is 118.7 kpc (4.6$'$).
 
We apply the mass-to-light versus colour relations (MLCRs), 
to estimate the stellar masses of AGC 111629 and AGC 748815. 
The MLCRs are taken from \cite{2015MNRAS.452.3209R}:
\begin{equation}
  \rm log(M_*/L_*)_g = 2.029 \times (g - r) - 0.984
\end{equation}  
This relation is based on the \cite{2003MNRAS.344.1000B} stellar population models and the \cite{2003PASP..115..763C} IMF. 
The galaxy colors are measured from the DECaLS g and r bands within the effective radius.

The (g - r) color is 0.42 mag for AGC 111629 and 0.21 mag for AGC 748815. 
The derived stellar masses are 7.46$\times$10$^8$ M$_{\odot}$ for AGC 111629 and 
1.0$\times$10$^8$ M$_{\odot}$ for AGC 748815.
The mass of AGC 111629 obtained from MLCRs is consistent with that reported in the MPA-JHU catalog (Table \ref{table:agc}). 
The stellar mass ratio of AGC 748815 to AGC 111629 is 0.13, suggesting that a minor merger may have occurred between them.

The occurrence of merger events in LSBGs has been investigated through both observations
(e.g., \citealt{2024ApJ...971..181C, 2025A&A...700A.165B, 2025ApJ...990..131C}) 
and simulations (e.g., \citealt{2018MNRAS.480L..18Z, 2023MNRAS.523.3991Z, 10.1088/1674-4527/ad5399, 2024MNRAS.533...93P, 2025arXiv250721231W}).

Resolved HI gas observations, which would reveal more extended structures, 
are needed to further investigate this interpretation. 

\subsection{Violent Activities in AGC 111629}

Violent activity in galaxies primarily includes starburst events, AGN, and supernova (SN) explosions, 
all of which can drive powerful winds, outflows, shock waves, and other forms of disruption.
M82 is a typical starburst galaxy, with a stellar mass of $\sim$1$\times$10$^{10}$ M$_{\odot}$ \citep{1992ApJ...395..126S}. 
Its star formation rate (SFR) is 10–13 M$_{\odot}$ yr$^{-1}$ \citep{1998PASJ...50..227S}, 
in contrast to the Milky Way's SFR of approximately 1.7 M$_{\odot}$ yr$^{-1}$.
We estimate the SFR of AGC 111629 based on its H$\alpha$ luminosity \citep{1998ARA&A..36..189K}, 
yielding a value of 0.077 M$_{\odot}$ yr$^{-1}$.
An independent estimate based on spectral energy distribution (SED) fitting from \citet{2024Univ...10..432C} 
gives a slightly lower SFR of 0.026 M$_{\odot}$ yr$^{-1}$.
Therefore, the SFR of AGC 111629 is far below the typical levels associated with starburst activity.

The central peanut-shaped structure in the [{\ion{O}{III}}]$\lambda$5007 image 
exhibits a different position angle (PA) from the overall galaxy and shows a significantly higher [{\ion{O}{III}}]$\lambda$5007/H$\beta$ flux ratio. 
These characteristics are consistent with features of an AGN ionization cone, 
which typically presents a distinct conical shape (e.g. NGC 5728, \citealt{2018ApJ...867..149D}).
Additionally, the H$\alpha$/H$\beta$ flux ratio within the peanut-shaped structure displays two low-value peaks, 
while both sides of the outer disk along the minor axis show significantly higher H$\alpha$/H$\beta$ ratios.
This may be explained if an AGN-driven wind has swept dust and metal-enriched gas into the outer regions.
We note both H$\alpha$ and [{\ion{O}{III}}]$\lambda$5007 do not show broad components in the center.

The stellar mass of AGC 111629 is below 10$^9$ M$_{\odot}$, 
making it plausible that it hosts a low-mass AGN. We estimate the black hole mass (M$_{\rm BH}$)
using the empirical relation between host galaxy stellar mass and black hole mass 
(logM$_{\rm BH}$ = 7.45 + 1.05 $\times$ log(M$_*$/10$^{11}$)) for dwarf galaxies from \cite{2015ApJ...813...82R}.
The expected M$_{\rm BH}$ is 5.1$\times$10$^5$ M$_{\odot}$ of AGC 111629.
The mean metallicity of the peanut-shaped structure is about 0.4Z$_{\odot}$.  
The metallicity and expected M$_{\rm BH}$ of AGC 111629 is consistent with the model data points in Figure\,\ref{BPT}.
Although the AGN component cannot be clearly identified through the spatially resolved BPT diagram based on [{\ion{N}{II}}], 
some spaxels lie in the composite region where is close to 100\% AGN model data point. 
The $|\rm EW|$ of [{\ion{O}{III}}]$\lambda$5007 shows one peak with a value of 6.5$\rm \AA$ in the peanut-shaped structure in Figure\,\ref{hab}. 
We note $|\rm EW|$ of [{\ion{O}{III}}]$\lambda$5007 of AGN can range from few $\rm \AA$ up to few hunderd $\rm \AA$ \citep{2005MNRAS.358.1043B, 2011MNRAS.415.1928C}.
There is a strong possibility that AGN exists in AGC 111629.

Superbubbles (SBs) can be produced by continuous energy input from supernova (SNe) explosions and may reach kiloparsec scales \citep{2003RMxAC..18..136T}.
The H$\alpha$/H$\beta$ flux ratio near the [{\ion{O}{III}}]$\lambda$5007 sub-peak shows a very extended region with unusually low values (Figure\,\ref{hab}), 
spanning approximately two kpc.
This low ratio could indicate either low dust attenuation or 
possibly linked to the Balmer self-absorption and scattering effects in nonspherically symmetric gas geometries
 \citep{2024arXiv240409015S, 2025arXiv250321896S}.

To the eastern of the sub-peak, the H$\alpha$/H$\beta$ flux ratio becomes very high, 
which may signal significant dust accumulation.
A possible explanation for this feature is that dust is being pushed outward by the expanding ejecta,
creating an asymmetric distribution of material around the region.

The rarity of detected supernovae in LSBGs \citep{2012A&A...538A..30Z} makes this potential SNe remnant signature particularly interesting. 
If confirmed, it would provide valuable insights into the role of stellar feedback in these diffuse systems. 
The effect of supernova feedback on LSBGs remains unknown, but if SNe can generate kiloparsec-scale structures like the one observed, 
they may play a significant role in shaping the interstellar medium (ISM) of LSBGs.
Future high-resolution observations and spectroscopic follow-ups could help determine whether 
the observed features are indeed linked to SN activity and constrain their influence on LSBG evolution.

\subsection{Radial Profile of Metallicity in LSBGs}
\cite{2024A&A...681A.100J} discovered a steep gradient from solar metallicity to subsolar values in the inner 20 kpc of Malin 1,
along with an almost flat gas-phase metallicity distribution in the outer disk, based on VLT/MUSE IFS observations.

\begin{figure*}
  \begin{center}
  \includegraphics[width=5.3in]{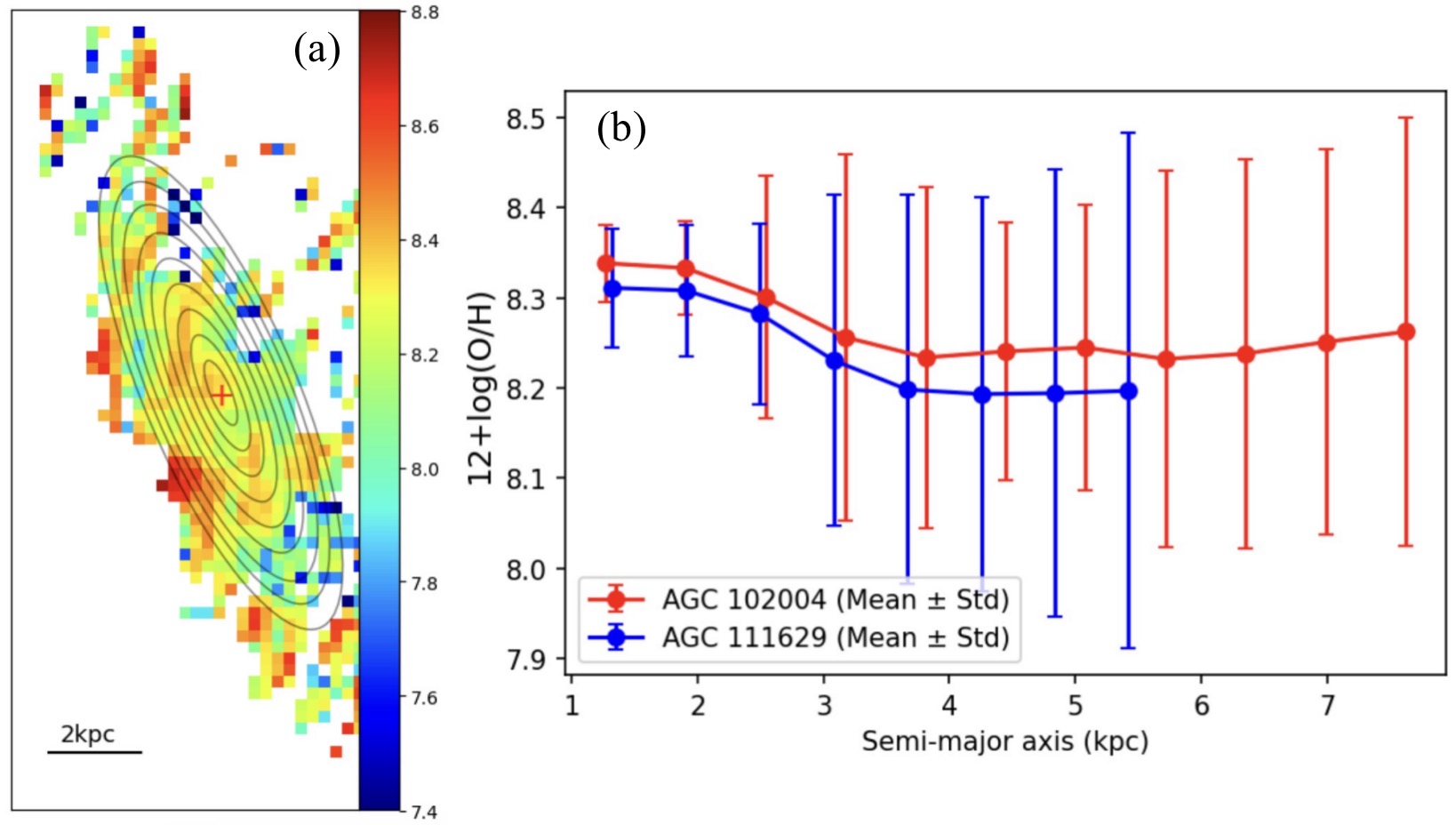}
  \end{center}
  \caption{Panel (a): Elliptical annuli overlaid on the gas-phase abundance map of 12+log(O/H), identical to Figure \ref{metal}(c).
Panel (b): Radial metallicity profiles of AGC 102004 (red dots) and AGC 111629 (blue dots), derived using the [N2] index \citep{2004MNRAS.348L..59P}. 
The profile of AGC 111629 is obtained from Panel (a), while that of AGC 102004 is adopted from Figure 8(c) in \cite{2024ApJ...971..181C}. 
In both cases, low-signal spaxels are included, and the H$\alpha$ morphology parameters from GALFIT are 
used to compute the mean 12+log(O/H) values within elliptical annuli.}
  \label{profile}
\end{figure*}

In Figure\,\ref{profile}, we present the radial metallicity profiles of 
AGC 102004 \citep{2024ApJ...971..181C} and AGC 111629. 
Both galaxies exhibit bright [\ion{N}{II}]$\lambda$6583 emission in their central regions 
and show overall low metallicity across their disks. 
The central regions (within 4 kpc) have slightly higher metallicities compared to their outer disks.
The radial profiles tend to flatten with increasing scatter in the outer regions, 
a trend consistent with that observed in Malin 1.

Both AGC 102004 and AGC 111629 exhibit low star formation rates (SFRs), 
and their flat metallicity distributions may indicate a stable star formation history. 
The H$\alpha$ disk of AGC 102004 displays a warped structure, 
while AGC 111629 exhibits an arch-like feature. 
\citet{2024ApJ...971..181C} analyzed a potential mini-merger in AGC 102004, 
and we discuss possible minor merger in AGC 111629 in Section 4.1.

Feedback from merger events, AGN activity, or supernova explosions 
may expel metal-rich gas or accrete metal-poor gas. 
Especially in low-mass galaxies, due to their shallow potential wells
feedback may be more important \citep{2018MNRAS.476..979P, 2020A&A...638A.123D, 2025ApJ...979...26S}.
The feedback processes may contribute to the overall low metallicity observed in LSBGs.
Additionally, cosmological gas accretion may also play a central role in setting 
the metallicity distribution \citep{2014A&ARv..22...71S}.

The metallicity differences between the inner and outer disks of our two ELSBGs 
are smaller than that observed in Malin 1 by \citet{2024A&A...681A.100J}. 
In Malin 1, the inner disk shows significantly higher metallicity than the outer disk, 
as illustrated in Figure 8 of \citet{2024A&A...681A.100J}. 
Giant LSBGs (GLSBGs) typically exhibit a central bulge-like structure \citep{2013JApA...34...19D}, whereas the ELSBGs from \citet{2023ApJ...948...96C} lack such bulges.
GLSBGs also present very different properties from normal LSBGs \citep{2023ApJ...959..105D}.
This indicats a different evolutionary path between GLSBGs and LSBGs. 

\section{Conclusions}
In this study, we present new IFS observations of ionized gas components
of an edge-on low surface brightness galaxy AGC 111629. 
Our primary findings are summarized as follows:

(1) AGC 111629 exhibits an irregular H$\alpha$ morphology and 
an arch-like structure in the extraplanar region that is absent in continuous stellar image. 
The irregular H$\alpha$ morphology may be related to a past merger event with its satellite
galaxy AGC 748815.

(2) A peanut-shaped structure at the center and a sub-peak in the southern disk are clearly visible
in the integrated [{\ion{O}{III}}]$\lambda$5007 image.
The position angle of the peanut-shaped structure differs from that of the entire galaxy.
Furthermore, the peanut-shaped structure shows a higher [{\ion{O}{III}}]$\lambda$5007/H$\beta$ flux ratio and 
inverse H$\alpha$/H$\beta$ flux radio distribution. 
The peanut-shaped structure may be associated with the central AGN, while the sub-peak likely corresponds 
to a superbubble created by supernova explosions in the southern disk.

(3) We derived the gas-phase metallicity, 12+log(O/H), using the [{\ion{N}{II}}]$\lambda$6583/H$\alpha$ diagnostic,
finding that AGC 111629 has low metallicity in the center. 
The low metallicity of AGC 111629 may be related to the feedback of the violent activities.
The radial variation of the metallicity for AGC 102004 and AGC 111629 is different from Malin 1,
which indicates the different formation paths of GLSBGs and normal LSBGs.

We present integral field spectroscopic (IFS) observations of two edge-on low surface brightness galaxies (ELSBGs)—AGC 102004 
\citep{2024ApJ...971..181C} and AGC 111629—from the sample in \cite{2023ApJ...948...96C}.
These studies utilize new IFS data to investigate diversity activities (e.g., AGN, supernova explosion, and merger events) in LSBGs. 
Our results contribute to a more comprehensive understanding of LSBGs from an edge-on perspective.

\section*{Acknowledgments}
We thank the anonymous referee for a number of very constructive comments.
We thank Prof. E. M. Di Teodoro help for resolving the problem we met using $^{3D}$BAROLO software.

This research uses data obtained through the Telescope Access Program (TAP), 
which has been funded by the TAP association, including Center for Astronomical Mega-Science CAS(CAMS), 
XMU, PKU, THU, USTC, NJU, YNU, and SYSU.

J.W. acknowledges National Key R\&D Program of China
(grant No. 2023YFA1607904) and the National Natural
Science Foundation of China (NSFC) grants Nos. 12033004,
12333002, and 12221003 and the science research grants
from the China Manned Space Project with Nos. 
CMS-CSST-2025-A10 and CMS-CSST-2025-A07. 
T.C. acknowledges the NSFC grants 12173045.
C.C. is supported by the NSFC, Nos. 11803044, 11933003,
and 12173045 and acknowledges the science research grants
from the China Manned Space Project with No. CMS-CSST-2021-A05. 
G.G. acknowledges the ANID BASAL projects ACE210002 and FB210003.
H.W. acknowledges the NSFC grant Nos. 11733006 and 12090041. 

This work is sponsored (in part) by the Chinese Academy of Sciences (CAS), through a grant to the
CAS South America Center for Astronomy (CASSACA), the NSFC grants 12090040 and 12090041, 
and the Strategic Priority Research Program of the
Chinese Academy of Sciences, grant No. XDB0550100.

\bibliographystyle{apj}
\bibstyle{thesisstyle}
\bibliography{main.bib}

\end{document}